\documentclass[aps,prb,preprint,superscriptaddress]{revtex4-2}

\usepackage{dcolumn}   
\usepackage{bm}        
\usepackage{amsmath}
\usepackage{amssymb}
\usepackage[utf8]{inputenc}
\usepackage{graphicx,siunitx}
\usepackage{amsmath}
\DeclareUnicodeCharacter{2212}{-}
\usepackage{booktabs}
\usepackage[svgnames]{xcolor}
\usepackage[export]{adjustbox}
\usepackage[percent]{overpic}
\usepackage[export]{adjustbox}
\usepackage{array}
\usepackage{caption}
\DeclareSIUnit\angstrom{\text {Å}}
\usepackage[linktocpage=true]{hyperref}
\usepackage{hyperref}
\usepackage{comment}
\usepackage{soul}
\usepackage{notes2bib}
\hypersetup{
	pdfnewwindow=true, 
	colorlinks=true,
	linkcolor=blue, 
	anchorcolor=blue,
	citecolor=blue, 
	filecolor=blue,
	menucolor=blue, 
	urlcolor=blue}

\usepackage{orcidlink}
\begin{document}
		\title{Pressure induced Electronic and Structural Transition in Ba$_2$NiTeO$_6$}
	
	\author{Bidisha Mukherjee\orcidlink{0000-0002-1485-7852}}
	\affiliation{Department of Physical Sciences, Indian Institute of Science Education and Research Kolkata, Mohanpur Campus, Mohanpur 741246, Nadia, West Bengal, India.}
    \affiliation{National Centre for High-Pressure Studies, Indian Institute of Science Education and Research Kolkata, Mohanpur Campus, Mohanpur 741246, Nadia, West Bengal, India.}
    
	\author{Supratik Mukherjee\orcidlink{0000-0002-6435-1374}}
	\affiliation{DRDO Industry Academia-Centre of Excellence (DIA-CoE), University of Hyderabad, Hyderabad, 500046, Telangana, India}
	\author{Mrinmay Sahu\orcidlink{0000-0003-4590-8904}}
	 \affiliation{Department of Physical Sciences, Indian Institute of Science Education and Research Kolkata, Mohanpur Campus, Mohanpur 741246, Nadia, West Bengal, India.}
     \affiliation{National Centre for High-Pressure Studies, Indian Institute of Science Education and Research Kolkata, Mohanpur Campus, Mohanpur 741246, Nadia, West Bengal, India.}
   
	\author{Bhagyashri Giri\orcidlink{0009-0002-9179-6771}}
	\affiliation{Department of Physical Sciences, Indian Institute of Science Education and Research Kolkata, Mohanpur Campus, Mohanpur 741246, Nadia, West Bengal, India.}
    \affiliation{National Centre for High-Pressure Studies, Indian Institute of Science Education and Research Kolkata, Mohanpur Campus, Mohanpur 741246, Nadia, West Bengal, India.}
    
	\author{A. C. Garcia-Castro\orcidlink{0000-0003-3379-4495}}
	\affiliation{School of Physics, Universidad Industrial de Santander, Carrera 27 Calle 09, Bucaramanga (Santander) 680002, Colombia}
	\author{Alfonso Mu\~noz\orcidlink{0000-0003-3347-6518}}
	\affiliation{Departamento de Física, MALTA Consolider Team, Universidad de La Laguna, San Cristóbal de La Laguna, Tenerife E-38200, Spain}
	\author{G. Vaitheeswaran\orcidlink{0000-0002-2320-7667}}
	\affiliation{School of Physics, University of Hyderabad, Prof. C. R. Rao Road, Gachibowli,
		Hyderabad 500046, Telangana, India}
        \email[]{vaithee@uohyd.ac.in}
	\author{Konstantin Glazyrin\orcidlink{0000-0002-5296-9265}}
	\affiliation{Photon Science, Deutsches Elektronen Synchrotron, 22607 Hamburg, Germany}
	\author{Goutam Dev Mukherjee\orcidlink{0000-0002-5388-743X}}
	\email [Corresponding author: ]{goutamdev@iiserkol.ac.in}
	\affiliation{Department of Physical Sciences, Indian Institute of Science Education and Research Kolkata, Mohanpur Campus, Mohanpur 741246, Nadia, West Bengal, India.}
    \affiliation{National Centre for High-Pressure Studies, Indian Institute of Science Education and Research Kolkata, Mohanpur Campus, Mohanpur 741246, Nadia, West Bengal, India.}
    
	\date{\today}
	
	\begin{abstract}
This study explores the pressure evolution of the double perovskite Ba$_2$NiTeO$_6$ by employing experimental and computational techniques. 
For the study of structural and vibrational properties, synchrotron X-ray diffraction (XRD) and micro-Raman spectroscopic experiments at high-pressures were carried out. As a complementary study, DFT simulations of the structural properties as a function of pressure were performed to support and explain the experimental findings. Furthermore, the electronic and magnetic properties as a function of pressure were investigated using DFT. Our study reveals a structural phase transition from a rhombohedral $R\bar{3}m$ to a monoclinic $C2/m$ phase at high pressure, accompanied by a significant increase in bulk modulus. Certain anomalies were observed in Raman mode frequencies at lower pressures of about 1 GPa, indicating changes in the electronic structure with a modification from direct to indirect bandgap in the sample. A minimum in the Raman mode full-width-half-maximum (FWHM) at about 11 GPa, coincides with an increase in ordering in the sample, indicated by a drop in the distortion index of Ni-O$_6$ octahedra as well as a discontinuity in the $c/a$ ratio. 
   
\end{abstract} 
\maketitle
	
\section{Introduction}
In condensed matter physics, high-pressure studies provide a powerful tool to probe and explore the stability of structural, electronic, and magnetic properties of materials under extreme conditions, offering critical insights into the behavior of complex oxides and their unique correlation among lattice, magnetic, and electronic degrees of freedom \cite{RevModPhys.90.015007,LIU2024299,Tse_2025,NCFO_me}. 
In this arena, double perovskites like Ba$_2$NiTeO$_6$ (BNTO) have garnered considerable interest due to their distinctive crystal structure and potential applications in electronic and magnetic devices \cite{BNTO_magnetic_LT,BNTO_dielectric_Magnetic}. 
The Ba$_2$$M^{2+}$Te$^{6+}$O$_6$ family ($M^{2+}$ = Co, Ni, Cu, or Zn) features a distinctive framework of face-sharing TeO$_6$ and $M$O$_6$ octahedra, interconnected by corner-sharing TeO$_6$ units \cite{Ba2CoTeO6,BNTO_buckledHoneycomb,BCuTOSurajitSaha,BZTOmoreiraPRM}. Among them, Ba$_2$NiTeO$_6$ (BNTO) crystallizes in a 12-layered rhombohedral structure ($R\bar{3}m$ space group) at room temperature, characterized by NiO$_6$–TeO$_6$–NiO$_6$ octahedral triplets. At 0.1~kHz, nanocrystalline BNTO possesses a large dielectric constant of value 96 \cite{BNTO_dielectric_Magnetic}. At lower temperatures, BNTO shows a magnetic transition with T$_N$ = 8.6~K~\cite{BNTO_magnetic_LT}. The application of pressure is known to alter atomic arrangements and, consequently, the physical properties of such materials, opening new avenues for technological applications \cite{HP_application1,HP_application2,HP_application3,HP_application4}. 
Notably, BaTeO$_3$, a single perovskite component of BNTO, exhibits abrupt changes in electrical transport and undergoes an irreversible structural transition at 12.7 GPa \cite{BaTeO3}. Additionally, the high-pressure-high-temperature synthesis process stabilizes BNTO in a hexagonal phase ($P6_3/mmc$) \cite{BNTO_HPSynthesis}. Nonetheless, despite its potential, no pressure-dependent studies on BNTO have been reported, leaving the potential new physics unexplored. 
With this motivation in mind, we explored, theoretically and experimentally, the pressure-induced structural, vibrational, and electronic properties of Ba$_2$NiTeO$_6$ in this study.

Ba$_2$ZnTeO$_6$ (BZTO), a non-magnetic analogue of BNTO, undergoes a ferroelastic structural phase transition at approximately 150~K, driven by a low-frequency phonon mode (E$_g$ mode) \cite{BZTOSurajitSaha, BZTOmoreiraPRM}. 
The frequency of this E$_g$ mode decreases progressively with temperature, reaching zero near the transition point. 
Similarly, BNTO also exhibits a low-frequency E$_g$ mode, the frequency of which decreases with lowering temperature but cannot reach the zero-frequency point. As such, the presence of a strong spin-phonon coupling in BNTO suppresses the occurrence of a structural phase transition \cite{BNTO_surajitSaha}. 
Our previous studies revealed that BZTO undergoes a pressure-induced $R\bar{3}m$ to $C2/m$ phase transition, also mediated by the E$_g$ mode \cite{BZTO_me}.

In this work, we aim to determine whether hydrostatic pressure can induce a structural phase transition in BNTO or if the strong spin-phonon coupling continues to inhibit such a transition, as observed in the low-temperature regime~\cite{BNTO_surajitSaha}.  
This paper presents a systematic investigation of the high-pressure behavior of BNTO, employing a combination of experimental techniques and theoretical simulations to elucidate the effects of elevated pressures on its lattice dynamics and electronic properties. Pressure-dependent synchrotron X-ray diffraction measurements are carried out up to about 45~GPa. 
The appearance of new Bragg peaks around 21~GPa indicates a structural phase transition. Further analysis of the XRD data shows a structural transition from rhombohedral $R\bar{3}m$ to monoclinic $C2/m$ phase. Pressure-dependent Raman spectra up to 40.3~GPa were recorded. Sudden slope changes of the peak centres and FWHM of the Raman modes with respect to pressure were detected at around 1~GPa, and 11~GPa. We have explained these anomalies found in the pressure evolution of Raman modes using both experimental results and DFT calculations. We have also discussed the interplay of the crystal symmetry with electronic and magnetic interactions in the sample within a stressed environment. Gaining insights into these connections can significantly advance the design of novel functional materials using transition metal oxides with tailored properties for emerging applications in electronics and spintronics.

\section{Experimental details}
High-purity powders of BaCO$_3$, NiO, and TeO$_2$ in suitable stoichiometric ratios were used for the solid-state synthesis of BNTO polycrystalline powder. These raw materials were mixed and ground with an agate mortar for 2 h. The corresponding reaction equation is as follows:
$$4 BaCO_3 + 2 NiO + 2 TeO_2 + O_2 = Ba_2NiTeO_6 + 2 CO_2$$
A light green powder was obtained after the grinding process, which was then heated at 1120℃ for 12 h in air. After the first heating process, an orange-yellow color compound was obtained. The powder was ground again for 1h in a mortar and heated in air at 1160℃ for 6h to get the final product. Using lab-XRD of Cu-K$_\alpha$ source, we confirmed that the single-phase pure sample.  
High-pressure Raman spectroscopic measurements were performed using a piston-cylinder type diamond anvil cell (DAC) having a culet of \SI{300}{\micro\metre} in diameter. A \SI{290}{\micro\metre} thick steel gasket was indented to a thickness of \SI{45}{\micro\metre}. At the centre of the indented region, a hole of diameter \SI{100}{\micro\metre} was drilled through the gasket using an electric discharge machine. The gasket was then placed on the lower diamond. A minimal amount of sample was loaded into the central hole, along with a 4:1 methanol-ethanol mixture as the pressure-transmitting medium (PTM), and ruby chips approximately \SI{5}{\micro\metre} in size were used for pressure calibration \cite{mao1986calibration}.

The Raman spectra of the sample were taken with a Monovista confocal micro-Raman system from S\&I GmbH equipped with a Cobolt Samba \SI{532}{\nano\metre} diode-pumped laser. A long-working-distance, infinitely corrected $20\times$ objective was used to focus the laser beam and to collect the scattered signal from the sample in the back-scattering geometry. The laser spot size on the sample surface was around \SI{4}{\micro\metre}. The collected light was dispersed using a grating with 1500 grooves/mm, and the spectra collected have a resolution of about~\SI{1.2}{\centi\metre}$^{-1}$. An edge filter for Rayleigh line rejection was used, which has a cut-off near ~\SI{80}{\centi\metre}$^{-1}$. 

High-pressure XRD data were taken at PETRA III, P02.2 beamline, Germany, using a monochromatic X-ray
of wavelength 0.2907~Å. For this experiment, the sample loading techniques are identical to the high-pressure Raman measurements, except that we used a Re gasket and neon gas as PTM. We calibrated the distance from the sample to the detector using the XRD pattern of a standard CeO$_2$ sample. Diffraction patterns were integrated to $2\theta$ versus intensity profile using DIOPTAS software \cite{dioptas}. To analyse the XRD data, GSAS \cite{toby2001expgui}, PCW \cite{kraus1996powder} EoSfit7 \cite{gonzalez2016eosfit7} and Vesta \cite{momma2008vesta} software were used.

\section{Computational Details}
We have performed density-functional \cite{PhysRev.136.B864,PhysRev.140.A1133} theoretical calculations as implemented in the
\textsc{vasp} code \cite{VASP,PhysRevB.59.1758} (version 5.4.4). In the projected-augmented waves, PAW \cite{PAW}, the approach was used to represent the valence and core electrons. The electronic configurations considered in the pseudo-potentials for the calculations were Ba: (5s$^2$5p$^6$6s$^2$, version 06Sep2000), Ni: (4s$^2$3d$^8$ version 06Sep2000), Te: (5s$^2$5p$^4$ version 08Apr2002), and O: (2s$^2$2p$^4$, version 08Apr2002). The exchange-correlation was represented within the generalized gradient approximation GGA-PBEsol \cite{PBESol} parametrization, and, due to strong exchange-correlation, we used the GGA$+U$ method based on Dudarev’s \cite {dudarev1998electron} method with a value of $U$ = 4.0 eV in the 3$d$ Ni states. The periodic solution of the crystal was represented by using Bloch states with a Monkhorst-Pack \cite{PhysRevB.13.5188} $k$-point mesh of 7$\times$7$\times$7, in the trigonal primitive cell, and 600 eV energy cut-off was utilized. Energy convergence below 0.00001 eV and the forces smaller that 0.001 eV~\AA$^{-1}$ and the stress tensor diagonal with a convergence criteria 0.1 GPa was implemented. Lattice-dynamical properties were obtained for the $\Gamma$-point using the direct-force constants approach at several pressures \cite{togo2023first}. The diagonalization of the dynamical matrix, which requires separate calculations of highly converged forces for small displacements, These calculations also allow the identification of the symmetry and eigenvectors of the vibrational modes. 
{
Crystal structure details from the XRD are given as input to initiate the DFT calculations. Finally, the atomic structure figures were elaborated with the \textsc{vesta} code \cite{vesta}.

\section{Results and discussions}

\subsection{High-pressure XRD study:}
The ambient XRD pattern could be indexed to the trigonal crystal structure in  $R\bar{3}m$ space group, having lattice constants $a = 5.7969(2)~\si{\angstrom}$, $c = 28.595(2)~\si{\angstrom}$ and volume, $V = 832.21(6)~\si{\angstrom}^3$. The obtained lattice parameters are in good agreement with the reported values in the literature \cite{BNTO_surajitSaha}. We have carried out the Rietveld refinement of the ambient XRD pattern using the atomic positions of BNTO, as reported by Badola et al. ~\cite{BNTO_surajitSaha}. The Rietveld refinement at ambient conditions is shown in FIG.~\ref{fig:Charecterization}, which shows an excellent fit with R$_p=1.27\%$ and R$_{wp}=2.19\%$ and the refined relative atomic positions are tabulated in TABLE~\ref{table:1}.
\begin{table}[ht!] 
	\begin{center}
		\begin{tabular}{ |c|c|c|c|c|c| } 
			\hline
			Atom & Wyckoff position & x/a & y/b & z/c\\
			\hline
			Ba & 6c & 0.0 & 0.0 & 0.1270(2)\\
			Ba & 6c & 0.0 & 0.0 & 0.2840(2)\\
			Ni & 6c & 0.0 & 0.0 & 0.4055\\
			Te & 3a & 0.0 & 0.0 & 0.00\\
			Te & 3b & 0.0 & 0.0 & 0.5\\
			O & 18h & 0.153 & -0.153 & 0.4593\\
			O & 18h & 0.177 & -0.177 & 0.6275\\
			\hline
		\end{tabular}\\
	\end{center}
	\caption{Relative atomic positions at ambient conditions after Rietveld refinement.}
	\label{table:1}
\end{table}
The unit cell of BNTO is a 12R (12 layers) type, which is formed with face sharing and corner sharing NiO$_{6}$ and TeO$_6$ octahedra, as shown in FIG.~\ref{fig:Charecterization}(b).

High-pressure XRD measurements were carried out up to about 45.2~GPa. The pressure evolution of XRD data of some selected pressure points is shown in FIG.~\ref{fig:EoS}(a).
All Bragg peaks shifted to the right side with increasing pressure. The first dashed lines from the left are guides to the new Bragg peak, which appeared at about 22.8~GPa, and the next two dashed lines show the splitting of a Bragg peak. The fourth line indicates the Bragg peak whose relative intensity increases with increasing pressure after 22.8~GPa. The XRD patterns could not be matched to the parent structure. A fresh indexing results in a monoclinic structure with space group $C2/m$. Lattice parameters obtained from LeBail refinement are $a= 9.599(3)$~\si{\angstrom}, $b= 5.578(2)$~\si{\angstrom}, $c= 9.700(3)$~\si{\angstrom}, $\beta= 108.78(2)$ and the corresponding unit cell volume is $491.7(2)$~\si{\angstrom}$^3$. To carry out the Rietveld refinement, the atomic positions from the high-pressure phase of its sister compound, BZTO were used. BZTO shows a similar structural transition to a monoclinic $C2/m$ phase from $R\bar{3}m$ phase on applying pressure. The Rietveld refinement at 22.8~GPa is shown in FIG.~\ref{fig:Charecterization}. The R$_p$ and R$_{wp}$ values are 4.08\% and 2.98\% respectively, which indicates an excellent fit. The obtained relative atomic positions after refinement are tabulated in TABLE~\ref{table:2}.
\begin{table}[ht!] 
	\begin{center}
		\begin{tabular}{ |c|c|c|c|c|c| } 
			\hline
			Atom & Wyckoff position & x/a & y/b & z/c\\
			\hline
			Ba & 4c & 0.133(1) & 0.0 & 0.3748(8)\\
			Ba & 4c & 0.291(2) & 0.0 & 0.8435(8)\\
			Ni & 4c & 0.902(4) & 0.5000 & 0.201(2)\\
			Te & 2a & 0.0 & 0.0 & 0.0\\
			Te & 2b & 0.0 & 0.5 & 0.5\\
			O & 4h & 0.1328 & 0.5 & 0.3998\\
			O & 8h & -0.1046 & 0.7283 & 0.3687\\
			O & 4h & 0.3177 & 0.5 & 0.8754\\
			O & 8h & 0.0608 & 0.6603 & 0.8999\\
			\hline
		\end{tabular}\\
	\end{center}
	\caption{Relative atomic positions of the high-pressure phase $C2/m$ after Rietveld refinement.}
	\label{table:2}
\end{table}

The pressure variation of lattice parameters is shown in FIG.~\ref{fig:EoS}(b). $R\bar{3}m$ phase has 60 atoms per unit cell, while the $C2/m$ phase has 40 atoms per unit cell. So, to compare the volume data before and after the structural transition, we have normalized the volume data of each phase with respect to their formula unit and shown it in FIG.~\ref{fig:EoS}(c). One can see a slight change in volume at the transition pressure, but the pressure evolution shows a clear discontinuity in slope. The experimental volume data of both phases were fitted using the third-order Birch-Murnaghan Equation of State (BM-EoS) \cite{PhysRev.71.809,doi:10.1073/pnas.30.9.244}. The bulk modulus (K$_0$) of the rhombohedral and monoclinic phases are 102(3)~GPa and 167(2)~GPa, respectively, and the first-order derivative of the bulk modulus (K$_p$) is 8.7(7) and 1.4(1), respectively.
To check the trigonality of the crystal structure as a function of pressure, we have plotted the ($c/a$) ratio and shown it in FIG.~\ref{fig:cByaRatio}. 
The ($c/a$) ratio is a crystallographic parameter that reflects the anisotropic strain in a trigonal unit cell. The value of ($c/a$) under ambient conditions is 4.9328(2), which indicates that the unit cell is highly anisotropic. With increasing pressure, the ($c/a$) value increases monotonically up to 12.5~GPa, followed by a sudden drop, and increases again with pressure thereafter. The increasing nature of the ($c/a$) ratio indicates that BNTO has stiffer bonding along the vertical direction than the basal plane. However, the sudden change in ($c/a$) value can reflect a change in the bonding environment or a change in distortion in the lattice, which can in turn influence the electronic or optical properties of the crystal. To investigate this in detail, we examined the distortion index (DI) and average bond length of NiO$_6$ octahedra as a function of pressure. The Distortion index of octahedra is calculated using the formula: $DI=\frac{1}{6}\sum \frac{(l_i-\bar{l})}{\bar{l}}$, where $l_i$ are the individual Ni--O bond lengths and $\bar{l}$ is the average bond length $\bar{l}=\frac{1}{6}\sum l_i$. The response of the distortion index of NiO$_6$ octahedra as a function of pressure is shown in FIG.~\ref{fig:octahedra}(a).

The distortion index has a maximum value at around 10.2~GPa. The pressure response of the average bond length also increases suddenly around the same pressure mentioned above, see FIG.~\ref{fig:octahedra}(b). 
An octahedron can distort to optimize the packing under higher pressures, which could lead to abrupt changes in its geometry. It can also affect the crystal field environment around the Ni$^{2+}$ ions or can be a precursor to lattice instabilities or the onset of a metastable phase. However, we can not conclude any of these possibilities from the XRD data alone.

\subsection{High-pressure Raman spectroscopic study:}
Raman spectroscopy enables us to probe changes in the vibrational modes of materials, providing insights into their structural dynamics under extreme conditions. According to group theoretical analysis, 16 modes (7A$_{1g}+9$E$_{g}$) are Raman active for $R\bar{3}m$ structure. The ambient Raman spectrum is in excellent agreement with those reported in the literature, having 16 Raman modes \cite{BNTO_surajitSaha}. 
The Raman modes are fitted to Lorentzian profiles and shown in FIG.~\ref{fig:AmbRaman}. Peak centres of the ambient Raman modes and their origin are tabulated in TABLE~\ref{table:3}. The pressure evolution of Raman spectra at some selected pressure points is shown in FIG.~\ref{fig:RamanVerticleTranslate}. Certain changes in Raman spectra are observed above 10 GPa, such as, merging of $N_{16}$ and $N_{17}$ modes; merging of $N_{13}$ and $N_{14}$ modes, merging of $N_{8}$ and $N_{9}$ modes, splitting of $N_{12}$ mode; disappearance of $N_5$ mode. However, XRD results show that the structural transition takes place above 21 GPa. One may attribute these changes in the Raman spectra above 10.2 GPa to the freezing of the PTM. Nonetheless, it should be pointed out that the $c/a$ ratio shows an anomaly at the same pressure, and the XRD data were taken with Ne as PTM, which remains hydrostatic. Therefore, one can also attribute these Raman spectra changes to the change in internal strain of the unit cell. Nevertheless, above 20 GPa, the Raman spectra become very broad and can be attributed to the structural transition.
A close inspection reveals certain inconsistencies in the pressure evolution of some Raman modes. 
In FIG.~\ref{fig:AllRaman}, we have shown the pressure evolution of a few selected Raman modes ($N_{10}$, $N_{12}$, $N_{13}$, $N_{14}$, $N_{16}$) till about 10 GPa. One finds a distinct slope change at about 1 GPa. In fact, the $N_{11}$ mode shows a softening till about 1.5 GPa. The above observations are quite intriguing because XRD analyses do not show any major anomaly at the same pressure other than a drop in the distortion index of $NiO_6$ octahedra and the average $Ni-O$ bond length. 
This indicates that the ordering of $NiO_6$ octahedra by the reduction of the $Ni-O$ bond length affects the Raman modes originating from the octahedra. 
The FWHM of a Raman mode represents the lifetime of the phonon, which is affected by anharmonic scattering with other phonons as well as other quasi-particles. 
In FIG.~\ref{fig:FWHM} we have shown the pressure variation of the FWHM of the above Raman modes. Interestingly, all of them show a sharp decrease till about 1 GPa. N$_{11}$ and N$_{16}$ mode-FWHM keep on decreasing till about 11 GPa and then increase. For other modes, it remains almost constant till about 11 GPa. However, the anomalous behavior around 1.1~GPa and 11~GPa in the absence of any structural transition is of particular interest.

\begin{table}[ht!]
	\addtolength{\tabcolsep}{-3pt}
	\hspace{-1cm}
	\begin{tabular}{ |c|c|c|c|c|c|c|} 
		\hline
		&  &  \multicolumn{4}{c|}{Raman shift (cm$^{-1}$)} & \\ \cline{3-6}
		
		Mode & Sym. & \multicolumn{2}{c|}{This work} & \multicolumn{2}{c|} {Previous work\cite{BNTO_surajitSaha}} &  Contribution\\ \cline{3-6}
		& &Expt. & DFT &Expt. & DFT  &  \\
		\hline
		N$_1$ & E$_g$ & 54 & 43 & 55 & 31 & Translation of Ba, Ni, Translation, O-rocking\\
		N$_2$ & A$_{1g}$ & 92.6 & 87 & 92 & 86 &Translation of Ba,Ni, O-scissoring\\
		N$_3$ & A$_{1g}$ & 109.4 & 104 & 109 & 104 & Translation of Ba,Ni, O-rotation\\
		N$_4$ & E$_g$ & 120.9 & 115 & 121 & 115 &  Translation of Ba, O-rotation\\
		N$_5$ & E$_g$ & 178.5 & 171 & 180 & 171 & Translation of Ba, Ni, Rotaion of TeO$_6$ \\
		N$_6$ & E$_g$ & 224.7 & 221 & 226 & 222 & Translation of Ni,O\\
		N$_7$ & A$_{1g}$ & 251.2 & 247 & 252 & - & -\\
		N$_8$ & A$_{1g}$ & 378 & 352 & 378 & 340 & Scissoring and rocking of O\\
		N$_9$ & E$_g$ & 390.7 & 382 & 391 & 382 & Translation of Ni, O-scissoring and wagging\\
		N$_{10}$ & -  & 409.4 & - & 409 & - & -\\
		N$_{11}$ & A$_{1g}$ & 476 & 440 & 482 & 440 & Translation of Ni, O-scissoring and wagging\\
		N$_{12}$ & E$_g$ & 581 & 573 & 581 & 572 & Translation of Ni,Symmetric \\&&&&&&and asymmetric stretching of O\\
		N$_{13}$ & E$_g$ & 612.2 & 597 & 613 & 597 & Symmetric and asymmetric stretching of O\\
		N$_{14}$ & A$_{1g}$ & 684 & 671 & 684 & 670 & Symmetric stretching of O\\
		N$_{15}$ & A$_{1g}$ & 734.5 & 709 & 746 & 704 & Ni translation, Symmetric stretching of O\\
		N$_{16}$ & - & 750.7 & & 752 & - & - \\
		\hline
	\end{tabular}\\
	\caption{Peak positions and assignments of BNTO Raman modes at ambient conditions of our experiment and reported literature. Sym. signifies symmetry.}
	\label{table:3}
\end{table}

The full-width half maximum, FWHM, of a Raman mode is inversely proportional to the lifetime of the corresponding phonon. An increase in anharmonic interactions due to phonon-phonon scattering can reduce the phonon lifetime, resulting in a broader FWHM \cite{FWHM_anharmonicity}. But up to about 1~GPa, the FWHM decreases with pressure. This increase in the lifetime of the phonon mode is contrary to the behavior of the unit cell, where the $c/a$ ratio and the octahedral distortion increase. So the question is whether any other order, either electronic or magnetic, develops that helps to increase the phonon lifetime. 
Therefore, we have carried out magnetic and electronic structure calculations using ab-initio DFT simulation. 
We considered one ferromagnetic (FM) and three antiferromagnetic (AFM) structures to understand the ground state configuration of the sample. 
The AFM3 structure (taking 120 atoms per unit cell) is identified as having the lowest ground-state energy; therefore, we proceed with the high-pressure calculations using this magnetic configuration, see FIG.~\ref{fig:magneticarrangement}. 
The calculations include Hubbard$+U$ (DFT$+U$) to improve the exchange correlation representation of the Ni-3$d$ states. 
The net magnetization for the AFM configuration is zero, as expected. The Ni site shows a magnetic moment close to 1.7~$\mu_B$ per atom, while the Ba, Te, and O atoms have negligible magnetic moments. Therefore, we have plotted the variation of Ni magnetic moments with respect to pressure. We observed a monotonous decrease in the magnetic moment with pressure. This indicates that the anomalies in Raman modes and their FWHM below 2~GPa are not due to any changes in magnetic moment. 
Next, we calculated the bandgap energy as a function of pressure (FIG.~\ref{fig:DFT_BG_MM}), which reveals that the sample is a direct band gap semiconductor at ambient conditions with a bandgap 2.591~eV and transforms to an indirect band gap semiconductor with bandgap 2.622~eV at about 2~GPa. A slight extrapolation of the indirect bandgap to lower pressures indicates that both direct and indirect bandgap cross at about 1 GPa. 
Any electron excitation in an indirect bandgap material will require the assistance of a phonon. The Raman experiments were carried out using the 532 nm laser, which has an energy value of 2.33 eV and hence creates a condition of Resonance Raman. Therefore, pressure-induced change in band structure introduces an electron-phonon coupling, and may result in the anomalous behaviour of Raman modes below 2~GPa. 
Now we want to understand the reason behind the minimum in the FWHM of Raman modes at about 10~GPa. High-pressure XRD investigations reveal two important anomalies in the structural parameters: 1. A sudden drop in $c/a$ ratio, 2. a minimum in DI of NiO$_6$ octahedra at 10~GPa. This shows an isostructural transition in the unit cell with a sudden increase in ordering, which results in a minimum in FWHM of phonon modes. In all possibilities, there may be a resonance in the sample that produces sharper Raman modes excited by 532~nm (2.35~eV) laser driven by the sudden increase in ordering in the lattice. 



\section{Conclusions}
This study investigated the high-pressure behaviour of BNTO, using both experimental and computational techniques. The findings reveal various connections between pressure, crystal symmetry, and electronic interactions in this transition-metal oxide. The key finding of this study is a structural phase transition from a rhombohedral $R\bar{3}m$ to a monoclinic $C2/m$ phase at high pressure, with the emergence of new Bragg peaks and changes in lattice parameters. This transition is accompanied by a significant increase in bulk modulus. Additionally, a sudden increase in the distortion index of NiO$_6$ octahedra aligns with changes in Ni-O bond lengths, suggesting modifications in the bonding environment under compression. Distinct slope variations in Raman mode frequencies, along with anomalies in the FWHM, further indicate the presence of both electronic and structural transitions at different pressure values. These observations provide valuable insights into the pressure-driven behaviour of Ba$_2$NiTeO$_6$, emphasizing its complex interplay between structural, electronic, and vibrational properties. These results highlight BNTO’s potential for applications in high-pressure sensors, spintronics, and catalysis. We hope this work advances the understanding of BNTO’s high-pressure evolution and highlights its promise for functional material design.

\section*{Acknowledgments}
The authors acknowledge the financial support from the Department of Science and Technology, Government of India, to perform the experiment under the DST-DESY project in P02.2 extreme beam line condition at PETRA III, Germany. BM acknowledges the UGC, Government of India, for the financial support to carry out the PhD work. A. Muñoz acknowledge the financial support provided by the Spanish Ministerio de Ciencia e Innovaci´on MCIN (DOI: 10.13039/501100011033) for project PID2022-138076NB-C44.
A. C. Garcia-Castro acknowledges the grant No. 4211 entitled “Búsqueda y estudio de nuevos compuestos antiperovskitas laminares con respuesta termoeléctrica mejorada para su uso en nuevas energías limpias” supported by Vicerrectoría de Investigaciones y Extensión, VIE--UIS. G.V  acknowledges the financial assistance from the Institute of Eminence, University of Hyderabad (UoHIoE-RC3-21-046) and the Param Rudra Super Computing facility at the IUAC New Delhi for providing the computational facility.

\section*{Data availability:}
Data produced and used in the development of this work will be made available from the author upon reasonable request.

\section*{author contribution statements}
{Bidisha Mukherjee and Goutam Dev Mukherjee conceived and supervised the project. Bidisha Mukherjee, Mrinmay Sahu, Bhagyashri Giri, and Konstantin Glazyrin carried out the experiments. Supratik Mukherjee performed the spin polarized electronic structure calculations under the supervision of G. Vaitheeswaran, A. Muñoz perfomed the high-pressure calculations and supervised the project, G Vaitheeswaran and A. C. Garcia-Castro planned the theory part and supervised the project.}

\newpage

\bibliographystyle{unsrt}
\bibliography{manuscript}

@article{doi:10.1073/pnas.30.9.244,
author = {F. D. Murnaghan },
title = {The Compressibility of Media under Extreme Pressures},
journal = {Proceedings of the National Academy of Sciences},
volume = {30},
number = {9},
pages = {244-247},
year = {1944},
doi = {10.1073/pnas.30.9.244},
URL = {https://www.pnas.org/doi/abs/10.1073/pnas.30.9.244},
eprint = {https://www.pnas.org/doi/pdf/10.1073/pnas.30.9.244}}

@article{PhysRev.71.809,
  title = {Finite Elastic Strain of Cubic Crystals},
  author = {Birch, Francis},
  journal = {Phys. Rev.},
  volume = {71},
  issue = {11},
  pages = {809--824},
  numpages = {0},
  year = {1947},
  month = {Jun},
  publisher = {American Physical Society},
  doi = {10.1103/PhysRev.71.809},
  url = {https://link.aps.org/doi/10.1103/PhysRev.71.809}
}

@article{vesta,
author = "Momma, Koichi and Izumi, Fujio",
title = "{{\it VESTA3} for three-dimensional visualization of crystal, volumetric and morphology data}",
journal = "Journal of Applied Crystallography",
year = "2011",
volume = "44",
number = "6",
pages = "1272--1276",
month = "Dec",
doi = {10.1107/S0021889811038970},
}

@article{PhysRevB.13.5188,
  title = {Special points for Brillouin-zone integrations},
  author = {Monkhorst, Hendrik J. and Pack, James D.},
  journal = {Phys. Rev. B},
  volume = {13},
  issue = {12},
  pages = {5188--5192},
  numpages = {0},
  year = {1976},
  month = {Jun},
  publisher = {American Physical Society},
  doi = {10.1103/PhysRevB.13.5188},
  url = {https://link.aps.org/doi/10.1103/PhysRevB.13.5188}
}

@article{PhysRevB.59.1758,
  title = {From ultrasoft pseudopotentials to the projector augmented-wave method},
  author = {Kresse, G. and Joubert, D.},
  journal = {Phys. Rev. B},
  volume = {59},
  issue = {3},
  pages = {1758--1775},
  numpages = {0},
  year = {1999},
  month = {Jan},
  publisher = {American Physical Society},
  doi = {10.1103/PhysRevB.59.1758},
  url = {https://link.aps.org/doi/10.1103/PhysRevB.59.1758}
}

@article{PhysRev.136.B864,
  title = {Inhomogeneous Electron Gas},
  author = {Hohenberg, P. and Kohn, W.},
  journal = {Phys. Rev.},
  volume = {136},
  issue = {3B},
  pages = {B864--B871},
  numpages = {0},
  year = {1964},
  month = {Nov},
  publisher = {American Physical Society},
  doi = {10.1103/PhysRev.136.B864},
  url = {https://link.aps.org/doi/10.1103/PhysRev.136.B864}
}

@article{PhysRev.140.A1133,
  title = {Self-Consistent Equations Including Exchange and Correlation Effects},
  author = {Kohn, W. and Sham, L. J.},
  journal = {Phys. Rev.},
  volume = {140},
  issue = {4A},
  pages = {A1133--A1138},
  numpages = {0},
  year = {1965},
  month = {Nov},
  publisher = {American Physical Society},
  doi = {10.1103/PhysRev.140.A1133},
  url = {https://link.aps.org/doi/10.1103/PhysRev.140.A1133}
}

@article{Tse_2025,
doi = {10.1088/1361-648X/ade83d},
url = {https://doi.org/10.1088/1361-648X/ade83d},
year = {2025},
month = {jul},
publisher = {IOP Publishing},
volume = {37},
number = {27},
pages = {273003},
author = {Tse, John S and Kuang, Huiyao and Yao, Yansun},
title = {Pressure-induced reactions in minerals: a condensed matter physics perspective},
journal = {Journal of Physics: Condensed Matter}
}

@incollection{LIU2024299,
title = {Chapter 11 - Condensed matter chemistry at high pressure},
editor = {Ruren Xu and Jihong Yu and Wenfu Yan},
booktitle = {Introduction to Condensed Matter Chemistry},
publisher = {Elsevier},
pages = {299-322},
year = {2024},
isbn = {978-0-443-16140-7},
doi = {https://doi.org/10.1016/B978-0-443-16140-7.00014-6},
url = {https://www.sciencedirect.com/science/article/pii/B9780443161407000146},
author = {Xiaoyang Liu and Yong Zhou and Peng Liu}
}

@article{RevModPhys.90.015007,
  title = {Solids, liquids, and gases under high pressure},
  author = {Mao, Ho-Kwang and Chen, Xiao-Jia and Ding, Yang and Li, Bing and Wang, Lin},
  journal = {Rev. Mod. Phys.},
  volume = {90},
  issue = {1},
  pages = {015007},
  numpages = {55},
  year = {2018},
  month = {Mar},
  publisher = {American Physical Society},
  doi = {10.1103/RevModPhys.90.015007},
  url = {https://link.aps.org/doi/10.1103/RevModPhys.90.015007}
}

@article{BNTO_magnetic_LT,
  title = {Magnetic ordering of the buckled honeycomb lattice antiferromagnet ${\rm{Ba}}_{2}{\rm{NiTeO}}_{6}$},
  author = {Asai, Shinichiro and Soda, Minoru and Kasatani, Kazuhiro and Ono, Toshio and Avdeev, Maxim and Masuda, Takatsugu},
  journal = {Phys. Rev. B},
  volume = {93},
  issue = {2},
  pages = {024412},
  numpages = {7},
  year = {2016},
  month = {Jan},
  publisher = {American Physical Society},
  doi = {10.1103/PhysRevB.93.024412},
  url = {https://link.aps.org/doi/10.1103/PhysRevB.93.024412}
}

@article{BNTO_dielectric_Magnetic,
    author = {Bijelić, Jelena and Tatar, Dalibor and Sahu, Manisha and Jagličić, Zvonko and Djerdj, Igor},
    title = {Size reduction-induced properties modifications of antiferromagnetic dielectric nanocrystalline $ \rm {B}a_2{N}i{M}{O}_6$ ({M = W, Te}) double perovskites},
    journal = {Oxford Open Materials Science},
    volume = {1},
    number = {1},
    pages = {itaa003},
    year = {2020},
    month = {10},
    issn = {2633-6979},
    doi = {10.1093/oxfmat/itaa003},
    url = {https://doi.org/10.1093/oxfmat/itaa003},
    eprint = {https://academic.oup.com/ooms/article-pdf/1/1/itaa003/45723594/itaa003.pdf},
}

@article{BaTeO3,
doi = {10.1209/0295-5075/98/66006},
url = {https://dx.doi.org/10.1209/0295-5075/98/66006},
year = {2012},
month = {jun},
publisher = {EDP Sciences, IOP Publishing and Società Italiana di Fisica},
volume = {98},
number = {6},
pages = {66006},
author = {Li, Yan and Han, Yonghao and Ma, Yanzhang and Zhu, Pinwen and Wang, Xin and Gao, Chunxiao},
title = {Pressure effects on grain boundary, electrical and vibrational properties of the polycrystalline $ \rm {B}a{T}e{O}_3$},
journal = {Europhysics Letters},
}

@article{BNTO_HPSynthesis,
title = {Pressure-induced phase transitions of hexagonal perovskite-like oxides},
journal = {Journal of Solid State Chemistry},
volume = {233},
pages = {492-496},
year = {2016},
issn = {0022-4596},
doi = {https://doi.org/10.1016/j.jssc.2015.11.028},
url = {https://www.sciencedirect.com/science/article/pii/S0022459615302607},
author = {Tomoya Aoba and Taneli Tiittanen and Hisayuki Suematsu and Maarit Karppinen}
}

@article{mao1986calibration,
  title={Calibration of the ruby pressure gauge to 800 kbar under quasi-hydrostatic conditions},
  author={Mao, HK and Xu, J-A and Bell, PM},
  journal={Journal of Geophysical Research: Solid Earth},
  volume={91},
  number={B5},
  pages={4673--4676},
  year={1986},
  publisher={Wiley Online Library}
}

@article{kraus1996powder,
  title={{POWDER CELL}--a program for the representation and manipulation of crystal structures and calculation of the resulting {X}-ray powder patterns},
  author={Kraus, Werner and Nolze, Gert},
  journal={Journal of applied Crystallography},
  volume={29},
  number={3},
  pages={301--303},
  year={1996},
  publisher={International Union of Crystallography}
}

@article{toby2001expgui,
  title={{EXPGUI}, a graphical user interface for {GSAS}},
  author={Toby, Brian H},
  journal={Journal of applied crystallography},
  volume={34},
  number={2},
  pages={210--213},
  year={2001},
  publisher={International Union of Crystallography}
}

@article{gonzalez2016eosfit7,
  title={{E}os{F}it7-{GUI}: a new graphical user interface for equation of state calculations, analyses and teaching},
  author={Gonzalez-Platas, Javier and Alvaro, Matteo and Nestola, Fabrizio and Angel, Ross},
  journal={Journal of Applied Crystallography},
  volume={49},
  number={4},
  pages={1377--1382},
  year={2016},
  publisher={International Union of Crystallography}
}

@article{momma2008vesta,
  title={{VESTA}: a three-dimensional visualization system for electronic and structural analysis},
  author={Momma, Koichi and Izumi, Fujio},
  journal={Journal of Applied crystallography},
  volume={41},
  number={3},
  pages={653--658},
  year={2008},
  publisher={International Union of Crystallography}
}

@article{dioptas,
  title={{DIOPTAS}: a program for reduction of two-dimensional {X}-ray diffraction data and data exploration},
  author={Prescher, Clemens and Prakapenka, Vitali B},
  journal={High Pressure Research},
  volume={35},
  number={3},
  pages={223--230},
  year={2015},
  publisher={Taylor \& Francis}
}

@article{BCuTOSurajitSaha,
	title={Emergence of a spin-liquid-like phase in the quantum spin ladder compound $ \rm {B}a_2{C}u{T}e{O}_6$ with chemical substitution},
	author={Badola, Shalini and Negi, Devesh and Joshi, Aprajita and Ali, Asif and Shankar Singh, Ravi and Saha, Surajit},
	journal={Physical Review B},
	volume={109},
	number={10},
	pages={L100405},
	year={2024},
	publisher={APS}
}

@article{BZTOmoreiraPRM,
	title={Raman and infrared spectroscopic investigations of a ferroelastic phase transition in $ \rm {B}a_2{Z}n{T}e{O}_6$ double perovskite},
	author={Moreira, Roberto L and Lobo, Ricardo PSM and Ramos, S{\'e}rgio LLM and Sebastian, Mailadil T and Matinaga, Franklin M and Righi, Ariete and Dias, Anderson},
	journal={Physical Review Materials},
	volume={2},
	number={5},
	pages={054406},
	year={2018},
	publisher={American Physical Society}
}

@article{BZTOSurajitSaha,
	title={Lattice dynamics across the ferroelastic phase transition in $ \rm {B}a_2{Z}n{T}e{O}_6$: a Raman and first-principles study},
	author={Badola, Shalini and Mukherjee, Supratik and Ghosh, B and Sunil, Greeshma and Vaitheeswaran, G and Garcia-Castro, AC and Saha, Surajit},
	journal={Physical Chemistry Chemical Physics},
	volume={24},
	number={34},
	pages={20152--20163},
	year={2022},
	publisher={Royal Society of Chemistry}
}

@article{Ba2CoTeO6,
  title={Synthesis and characterisation of the vibrational and electrical properties of antiferromagnetic $ \rm 6{L}-{B}a_2{C}o{T}e{O}_6$ ceramics},
  author={Aziz, Anees A and Mercone, Silvana and Lobo, Ricardo PSM and Dias, Anderson and Moreira, Roberto L and Bell, Anthony MT and Eccleston, Roger and Feteira, Antonio},
  journal={Dalton Transactions},
  volume={48},
  number={29},
  pages={11112--11121},
  year={2019},
  publisher={Royal Society of Chemistry}
}

@article{BNTO_buckledHoneycomb,
  title = {Spin dynamics in the stripe-ordered buckled honeycomb lattice antiferromagnet $ \rm {B}a_2{N}i{T}e{O}_6$},
  author = {Asai, Shinichiro and Soda, Minoru and Kasatani, Kazuhiro and Ono, Toshio and Garlea, V. Ovidiu and Winn, Barry and Masuda, Takatsugu},
  journal = {Phys. Rev. B},
  volume = {96},
  issue = {10},
  pages = {104414},
  numpages = {6},
  year = {2017},
  month = {Sep},
  publisher = {American Physical Society},
  doi = {10.1103/PhysRevB.96.104414}
}

@article{BNTO_surajitSaha,
  title={Spin-phonon coupling suppressing the structural transition in perovskite-like oxide},
  author={Badola, Shalini and Mukherjee, Supratik and Sunil, Greeshma and Ghosh, B and Negi, Devesh and Vaitheeswaran, G and Garcia-Castro, AC and Saha, Surajit},
  journal={Physical Review B},
  volume={109},
  number={6},
  pages={L060104},
  year={2024},
  publisher={APS}
}

@article{BZTO_me,
  title={Soft mode induced structural phase transition in $ \rm {B}a_2{Z}n{T}e{O}_6$ at high pressure},
  author={Mukherjee, Bidisha and Adhikari, Surajit and Sahu, Mrinmay and Mishra, Asish Kumar and Giri, Bhagyashri and Johari, Priya and Glazyrin, Konstantin and Mukherjee, Goutam Dev},
  journal={Physical Review B},
  volume={111},
  number={17},
  pages={174113},
  year={2025},
  publisher={APS}
}

@article{HP_application1,
  title={High pressure crystal structures of orthovanadates and their properties},
  author={Errandonea, Daniel},
  journal={Journal of Applied Physics},
  volume={128},
  number={4},
  year={2020},
  publisher={AIP Publishing}
}

@article{HP_application2,
author ="Lü, Xujie and Yang, Wenge and Jia, Quanxi and Xu, Hongwu",
title  ="Pressure-induced dramatic changes in organic–inorganic halide perovskites",
journal  ="Chem. Sci.",
year  ="2017",
volume  ="8",
issue  ="10",
pages  ="6764-6776",
publisher  ="The Royal Society of Chemistry",
doi  ="10.1039/C7SC01845B"}

@article{HP_application3,
    author = {Navrotsky, Alexandra},
    title = {Pressure-induced structural changes cause large enhancement of photoluminescence in halide perovskites: a quantitative relationship},
    journal = {National Science Review},
    volume = {8},
    number = {9},
    pages = {nwab041},
    year = {2021},
    month = {03},
    issn = {2095-5138},
    doi = {10.1093/nsr/nwab041}
}

@article{HP_application4,
  title={Structural transition and emission enhancement in vacancy-ordered halide double perovskite $ \rm {C}s_2{T}e{C}l_6$ under pressure},
  author={Mukherjee, Suvashree and Samanta, Debabrata and Mukherjee, Bidisha and Glazyrin, Konstantin and Mukherjee, Goutam Dev},
  journal={Applied Physics Letters},
  volume={126},
  number={14},
  year={2025},
  publisher={AIP Publishing}
}

@article{NCFO_me,
  title={Pressure-induced softening in bulk modulus due to magnetoelastic coupling in $ \rm {N}d_2{C}o{F}e{O}_6$ double perovskite},
  author={Mukherjee, Bidisha and Sahu, Mrinmay and Samanta, Debabrata and Ghosh, Bishnupada and Joseph, Boby and Dev Mukherjee, Goutam},
  journal={Journal of Applied Physics},
  volume={136},
  number={9},
  year={2024},
  publisher={AIP Publishing}
}

@article{FWHM_anharmonicity,
  title={Anharmonicity},
  author={Cowley, RA},
  journal={J. Phys.(Paris)},
  volume={26},
  number={3},
  pages={659--664},
  year={1965}
}

@article{VASP,
	title = {Efficient iterative schemes for ab initio total-energy calculations using a plane-wave basis set},
	author = {Kresse, G. and Furthm\"uller, J.},
	journal = {Phys. Rev. B},
	volume = {54},
	issue = {16},
	pages = {11169--11186},
	numpages = {0},
	year = {1996},
	month = {Oct},
	publisher = {American Physical Society},
	doi = {10.1103/PhysRevB.54.11169},
	url = {https://link.aps.org/doi/10.1103/PhysRevB.54.11169}
}

@article{PAW,
	title = {Projector augmented-wave method},
	author = {Bl\"ochl, P. E.},
	journal = {Phys. Rev. B},
	volume = {50},
	issue = {24},
	pages = {17953--17979},
	numpages = {0},
	year = {1994},
	month = {Dec},
	publisher = {American Physical Society},
	doi = {10.1103/PhysRevB.50.17953},
	url = {https://link.aps.org/doi/10.1103/PhysRevB.50.17953}
}

@article{PBESol,
	title = {Restoring the Density-Gradient Expansion for Exchange in Solids and Surfaces},
	author = {Perdew, John P. and Ruzsinszky, Adrienn and Csonka, G\'abor I. and Vydrov, Oleg A. and Scuseria, Gustavo E. and Constantin, Lucian A. and Zhou, Xiaolan and Burke, Kieron},
	journal = {Phys. Rev. Lett.},
	volume = {100},
	issue = {13},
	pages = {136406},
	numpages = {4},
	year = {2008},
	month = {Apr},
	publisher = {American Physical Society},
	doi = {10.1103/PhysRevLett.100.136406},
	url = {https://link.aps.org/doi/10.1103/PhysRevLett.100.136406}
}

@article{dudarev1998electron,
	title={{Electron-energy-loss spectra and the structural stability of nickel oxide: An LSDA+ U study}},
	author={Dudarev, Sergei L and Botton, Gianluigi A and Savrasov, Sergey Y and Humphreys, CJ and Sutton, Adrian P},
	journal={{Physical Review B}},
	volume={57},
	number={3},
	pages={1505},
	year={1998},
	publisher={APS}
}

@article{togo2023first,
  title={First-principles phonon calculations with phonopy and phono3py},
  author={Togo, Atsushi},
  journal={Journal of the Physical Society of Japan},
  volume={92},
  number={1},
  pages={012001},
  year={2023},
  publisher={The Physical Society of Japan}
}

\newpage 

\section{Figures}		
\begin{figure}[h]			
		\includegraphics[width=0.5\linewidth]{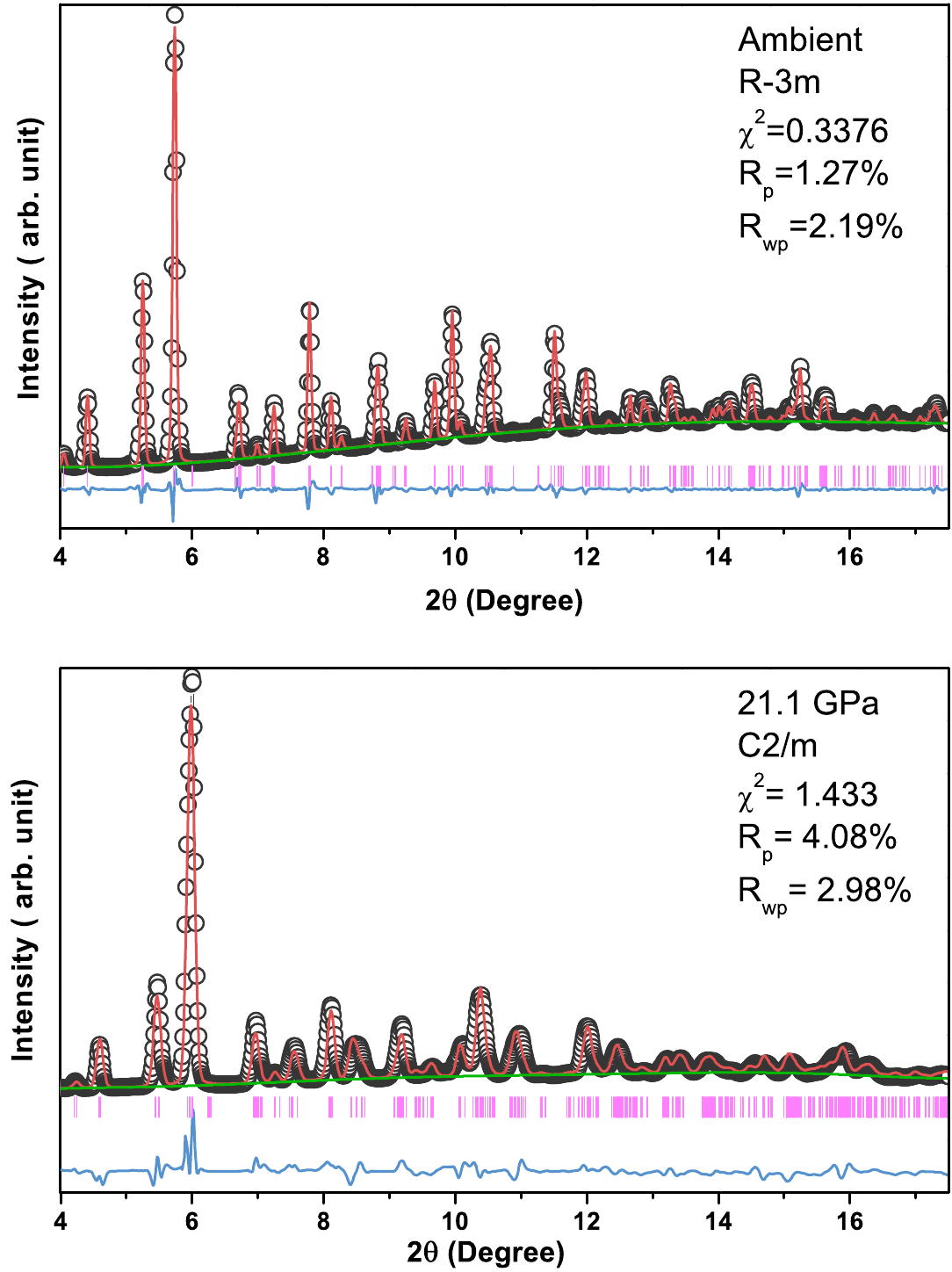}
		\includegraphics[width=0.34\linewidth]{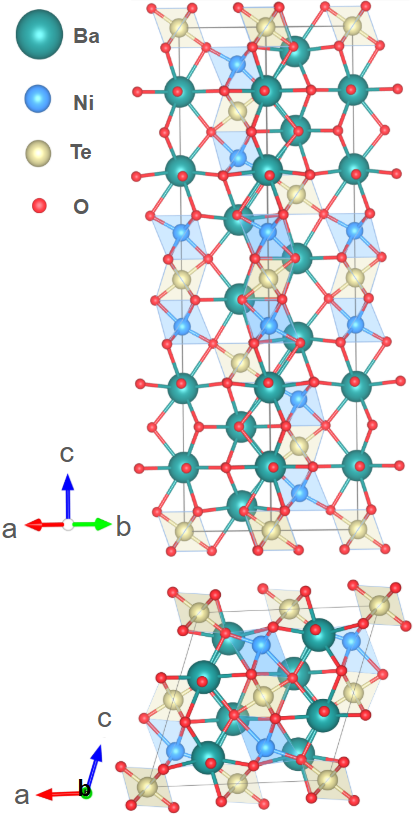}
		\caption{(Color online) \textbf{Left Panel: }{}Rietveld refinement of the XRD pattern at ambient conditions (top) and 21.1~GPa (bottom). Black circles represent experimental data and its Rietveld fit is plotted with a red line. The green line is the background and the difference between experimental and calculated data is shown with the blue line. The magenta vertical lines indicate the Bragg peaks of the sample. \textbf{Right Panel: }The representation of unit cell crystal structure of both the phases.}
	\label{fig:Charecterization}
\end{figure}
 \begin{figure}
     \centering
 	\includegraphics[width=0.9\linewidth]{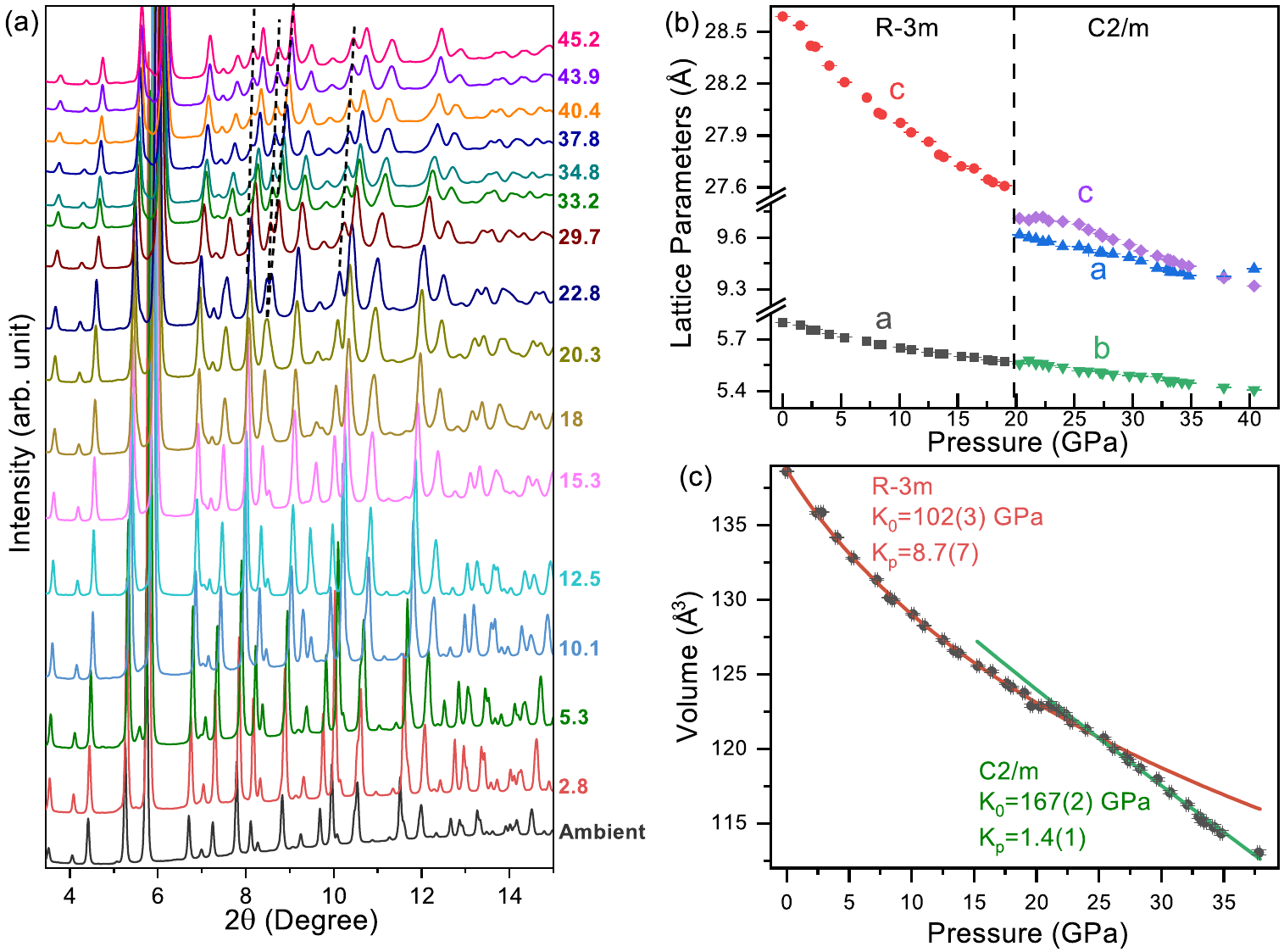}    
 	\caption{(Color online) (a) XRD pattern of BNTO taken at different pressures. Pressures (in GPa) are indicated alongside the respective data. The black arrows indicate the appearance of new peaks. (b) Pressure evolution of lattice parameters. (c) Black solid circles represent the volume data, and the red and olive lines are 3$^{rd}$ order BM-EoS fit to the experimental data.}
 	\label{fig:EoS}
 \end{figure}
 \begin{figure}
	\centering
	\includegraphics[width=0.50\linewidth]{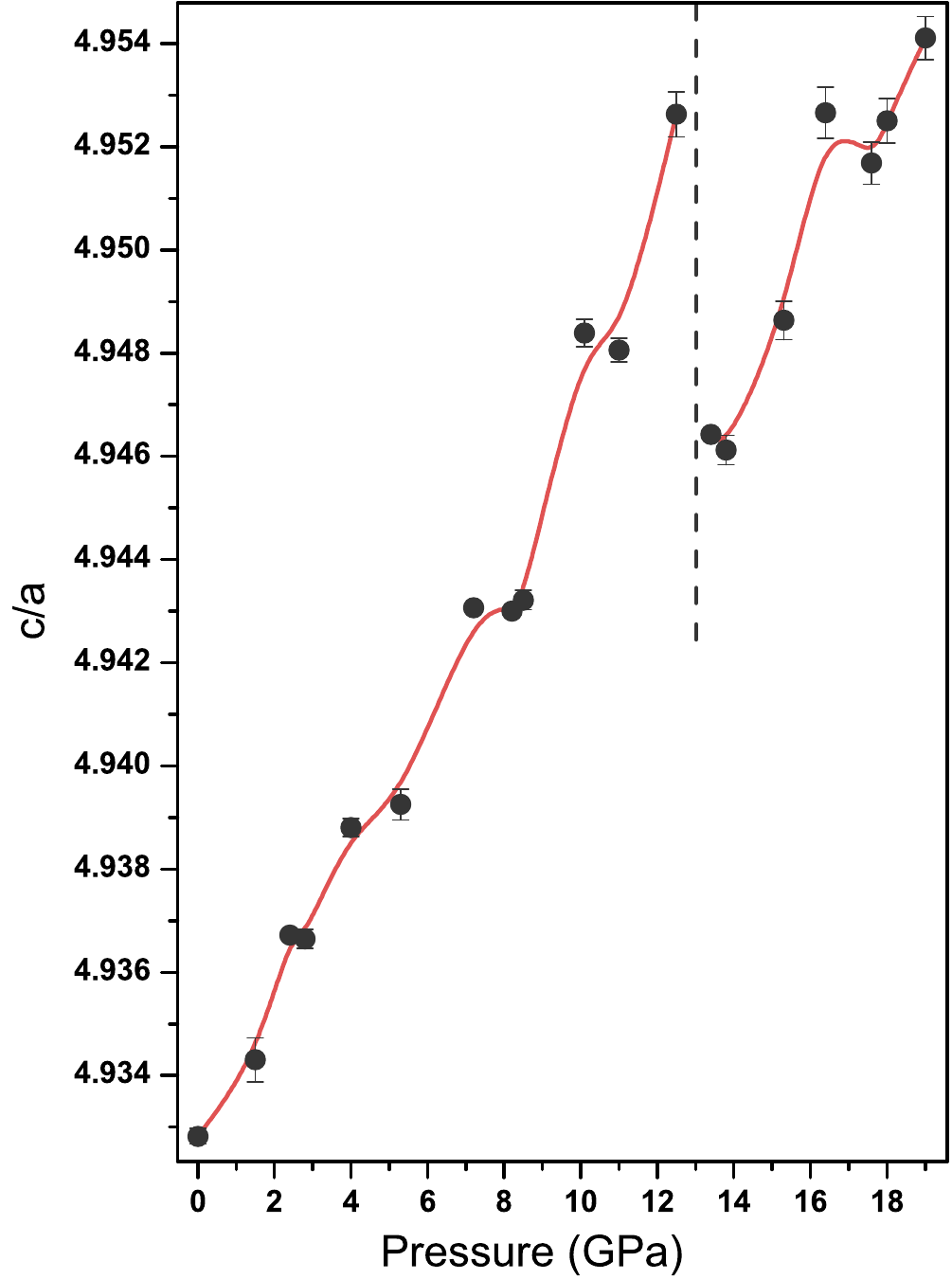}
	\caption{(Color online) The ratio between lattice parameters $c/a$ is plotted with solid circles with error bars. The red line is a spline fit to the data. A sudden drop in the $c/a$ ratio is observed at 12.5~GPa indicated by the dashed vertical line.}
	\label{fig:cByaRatio}
\end{figure} 
 \begin{figure}
 \centering
        \includegraphics[width=1.0\linewidth]{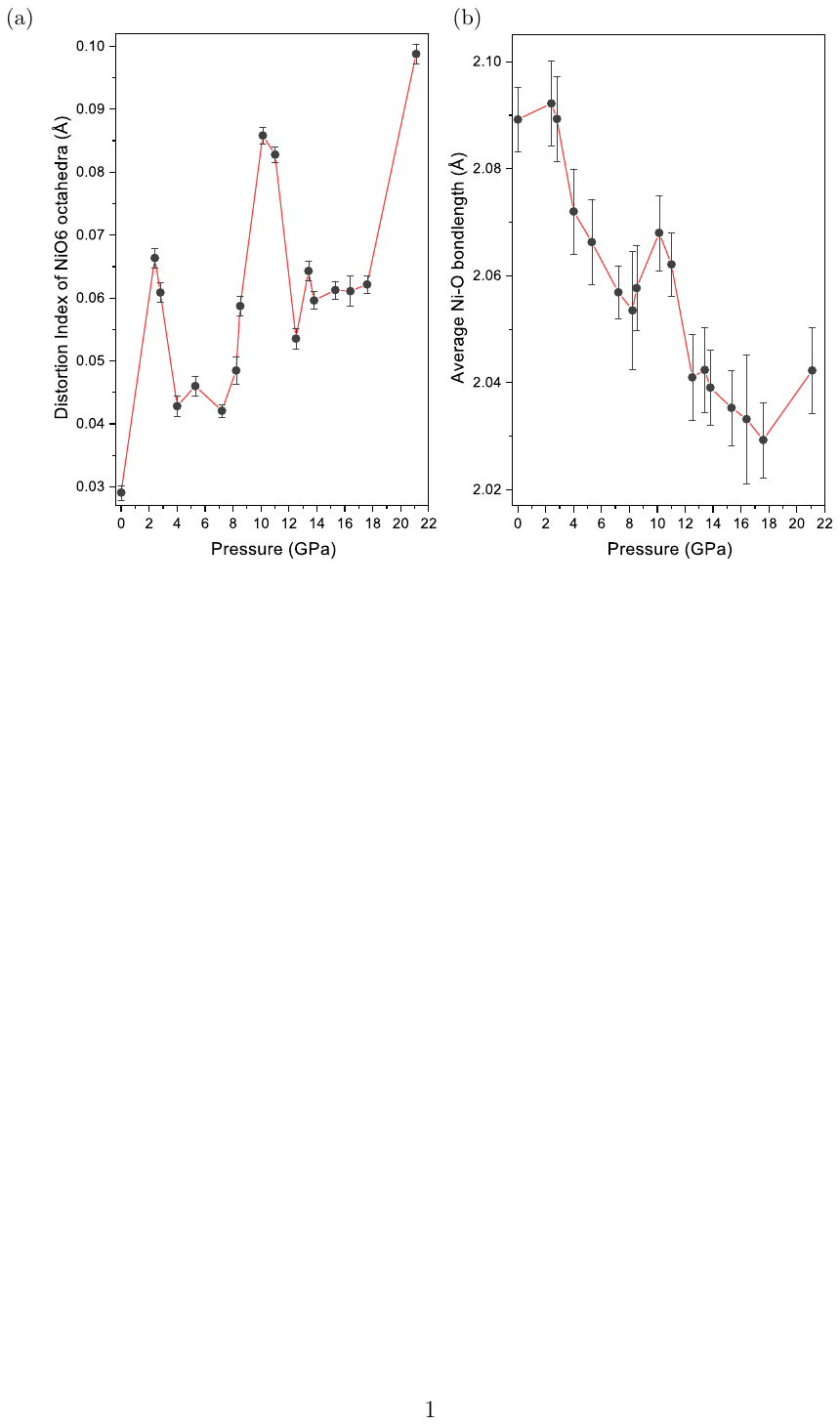}
        \vspace{-10cm}
	\caption{(Color online) Pressure evolution of (a) distortion index of NiO6 octahedra and (b) Ni--O average bond length. }
	\label{fig:octahedra}
\end{figure} 
\begin{figure}
	\centering
	\includegraphics[width=0.60\linewidth]{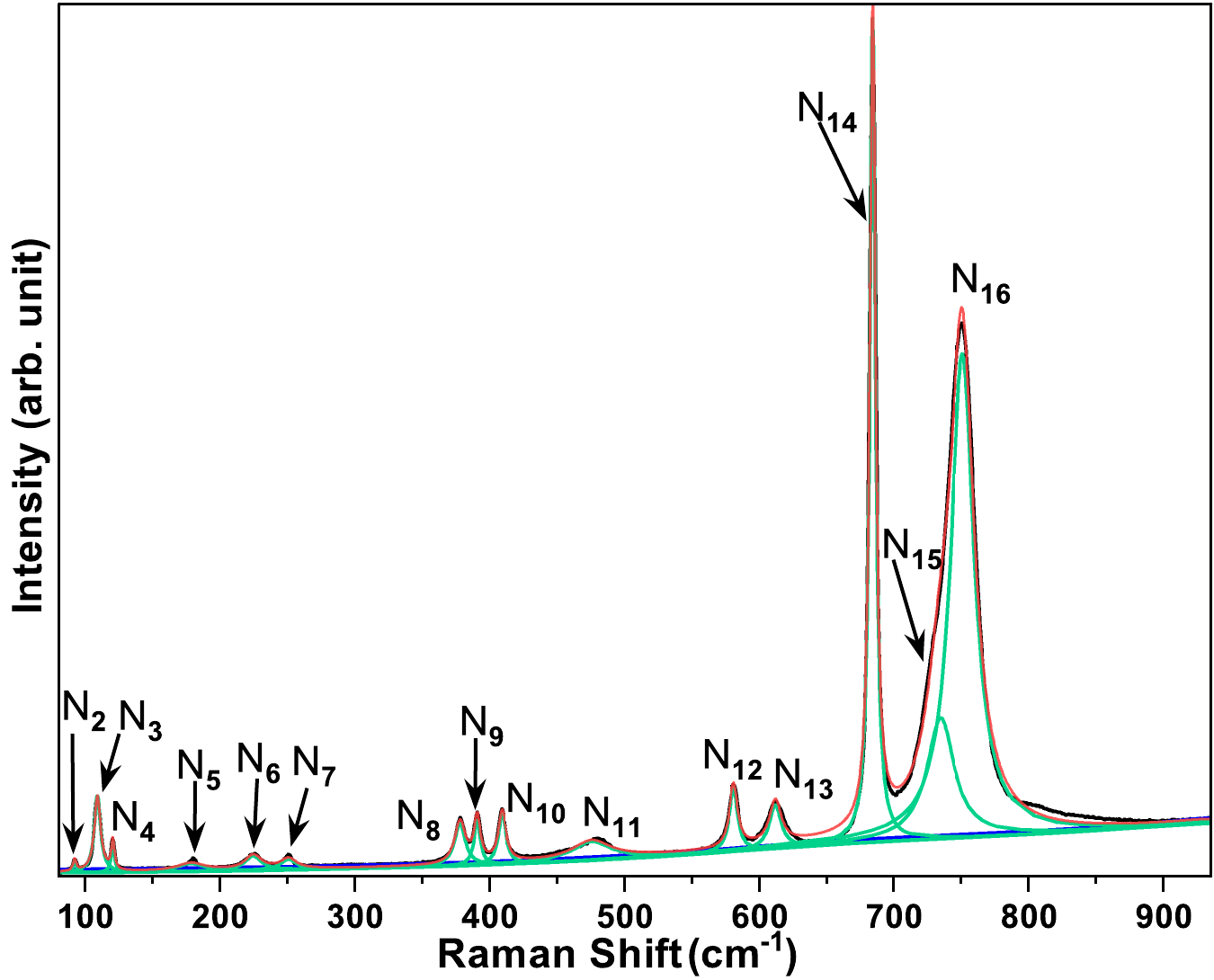}
	\caption{(Color online) Raman spectra of BNTO collected at ambient conditions. The background corrected spectrum is fitted using the Lorentzian profile.}
	\label{fig:AmbRaman}
\end{figure} 
\begin{figure}
	\centering
    \includegraphics[width=1.0\linewidth]{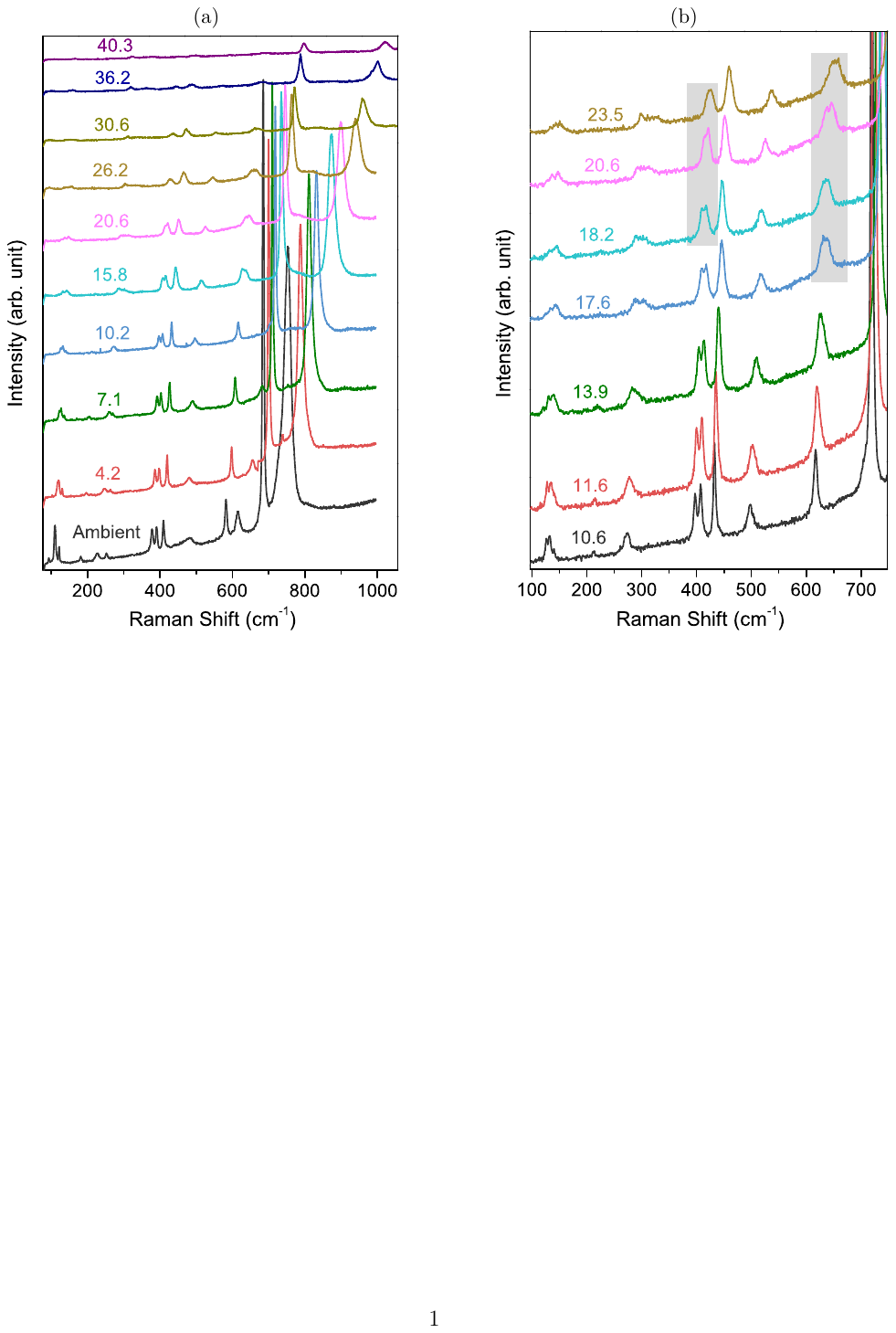}
        \vspace{-10cm}
		\caption{(Color online) (a) Pressure evolution of Raman spectra for a few selected pressure points. Pressure values (in GPa) are indicated alongside. (b) A zoomed-up version around the pressure of structural transition with grey highlights indicating a merging and a splitting of Raman peaks. }
		\label{fig:RamanVerticleTranslate}
\end{figure}
 \begin{figure}
 	\centering
 	\includegraphics[trim=3cm 10cm 2cm 0cm,clip]{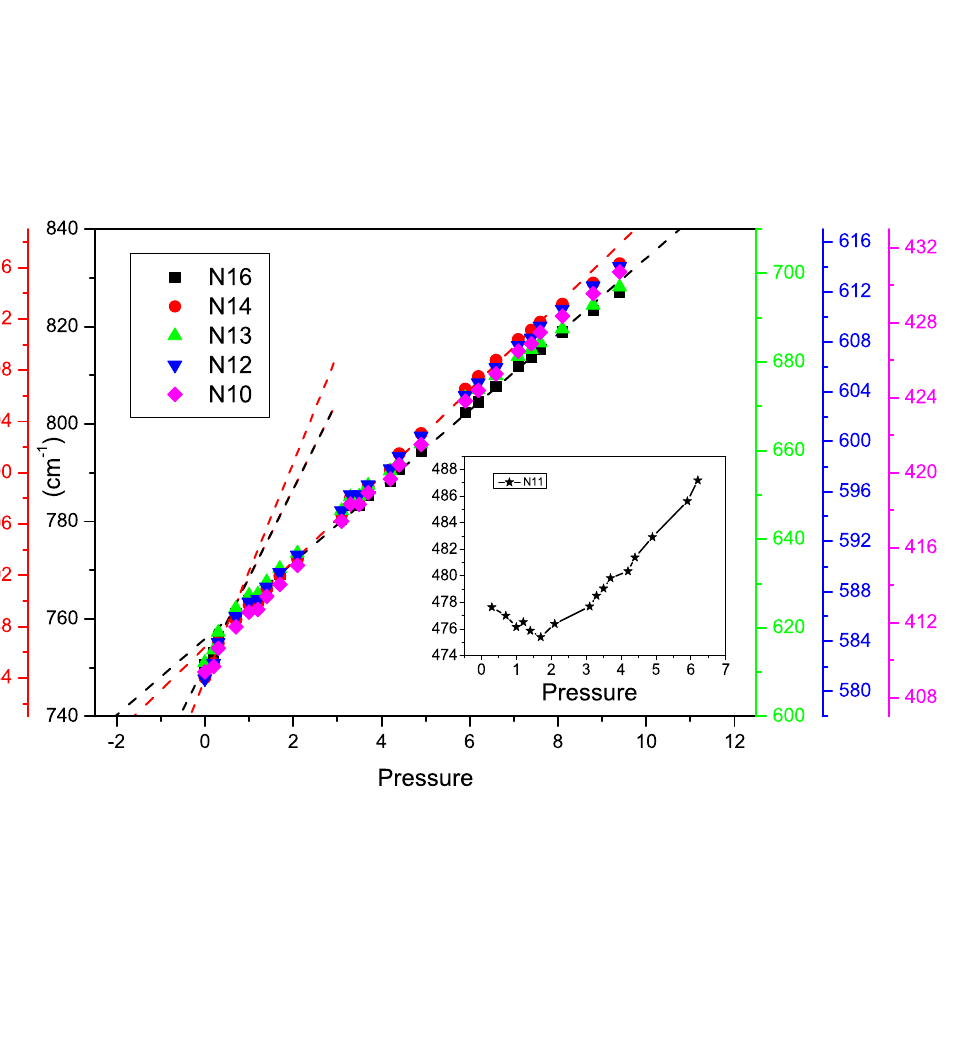}
 	
 	\caption{(Color online) Pressure evolution of peak positions of selected Raman modes showing a slope change at about 1 GPa. The inset shows softening of N$_{11}$ Raman mode at the low-pressure region. }
 	\label{fig:AllRaman}
 \end{figure} 

\begin{figure}
	\centering
    \includegraphics[trim=3cm 10cm 0cm 0cm,clip]{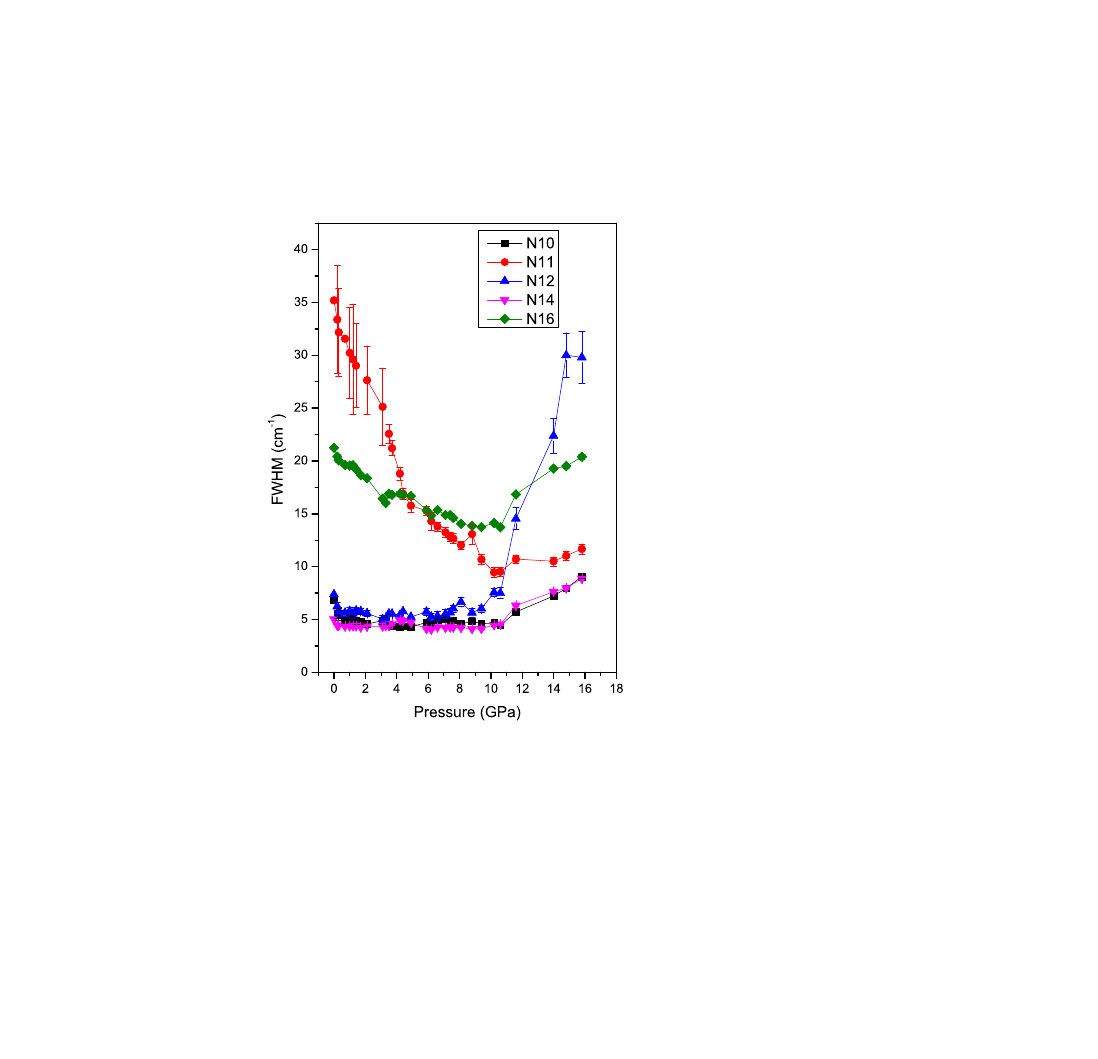}
 	
	\caption{(Color online) Evolution of the full-width half maximum, FWHM, of the N$_{16}$, N$_{14}$, N$_{12}$, N$_{11}$, and N$_{10}$ modes as a function of the applied pressure.}
	\label{fig:FWHM}
\end{figure}
 \begin{figure}
     \centering
 	\includegraphics[trim=2cm 10cm 0cm 0cm,clip]{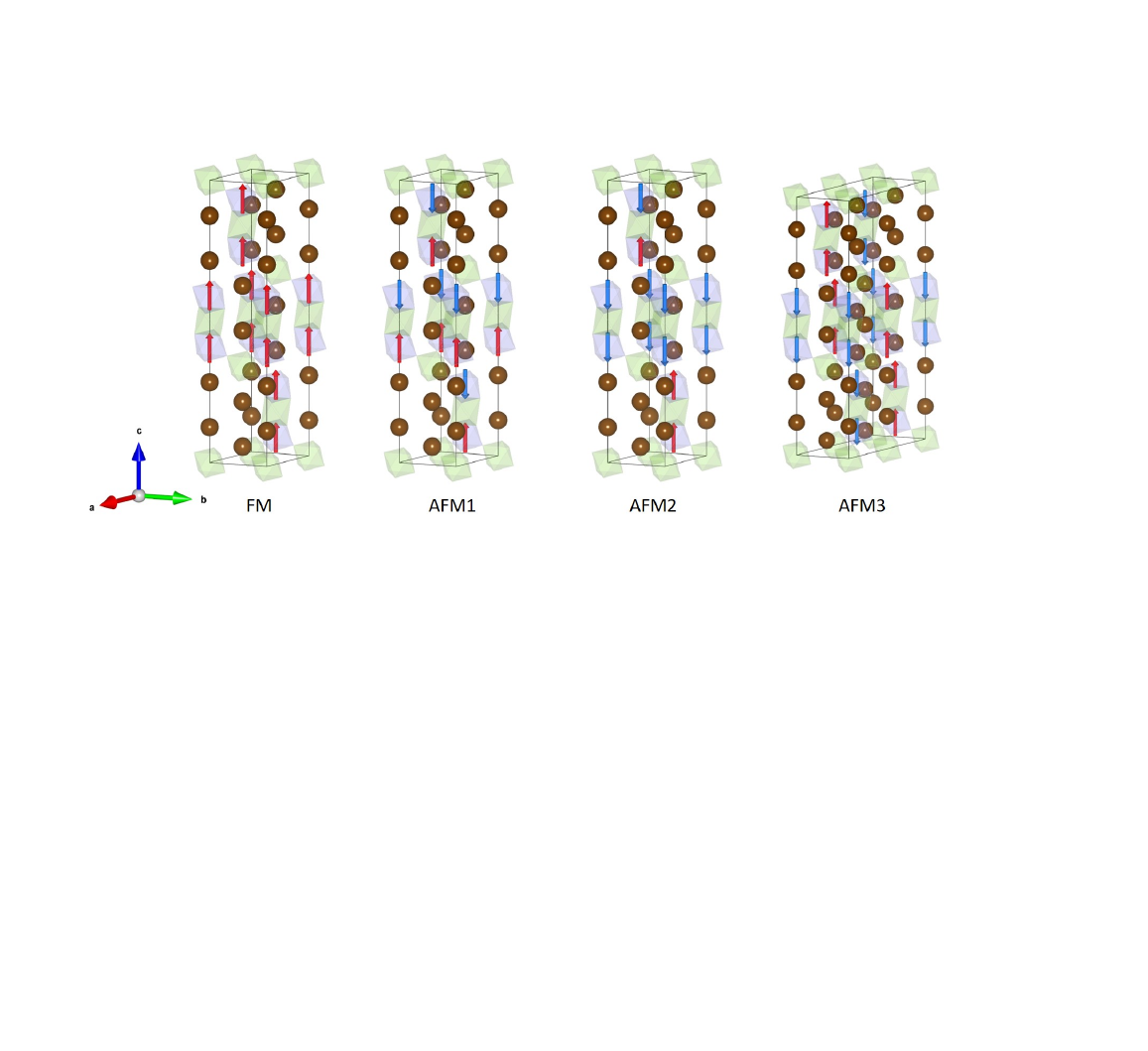}

     \caption{(Color online) Four different types of magnetic configurations are considered, namely ferromagnetic (FM) and 3 different types of antiferromagnetic (AFM) (left to right). In the AFM1 and AFM2, the magnetic propagation q-vector is (0,0,0), whereas in the AFM3 case, the q-vector is equivalent to  (1/2,0,0) with reference to the hexagonal unit cell.}
     \label{fig:magneticarrangement}
\end{figure}
\begin{figure}
\centering
\includegraphics[trim=2cm 10cm 0cm 0cm,clip]{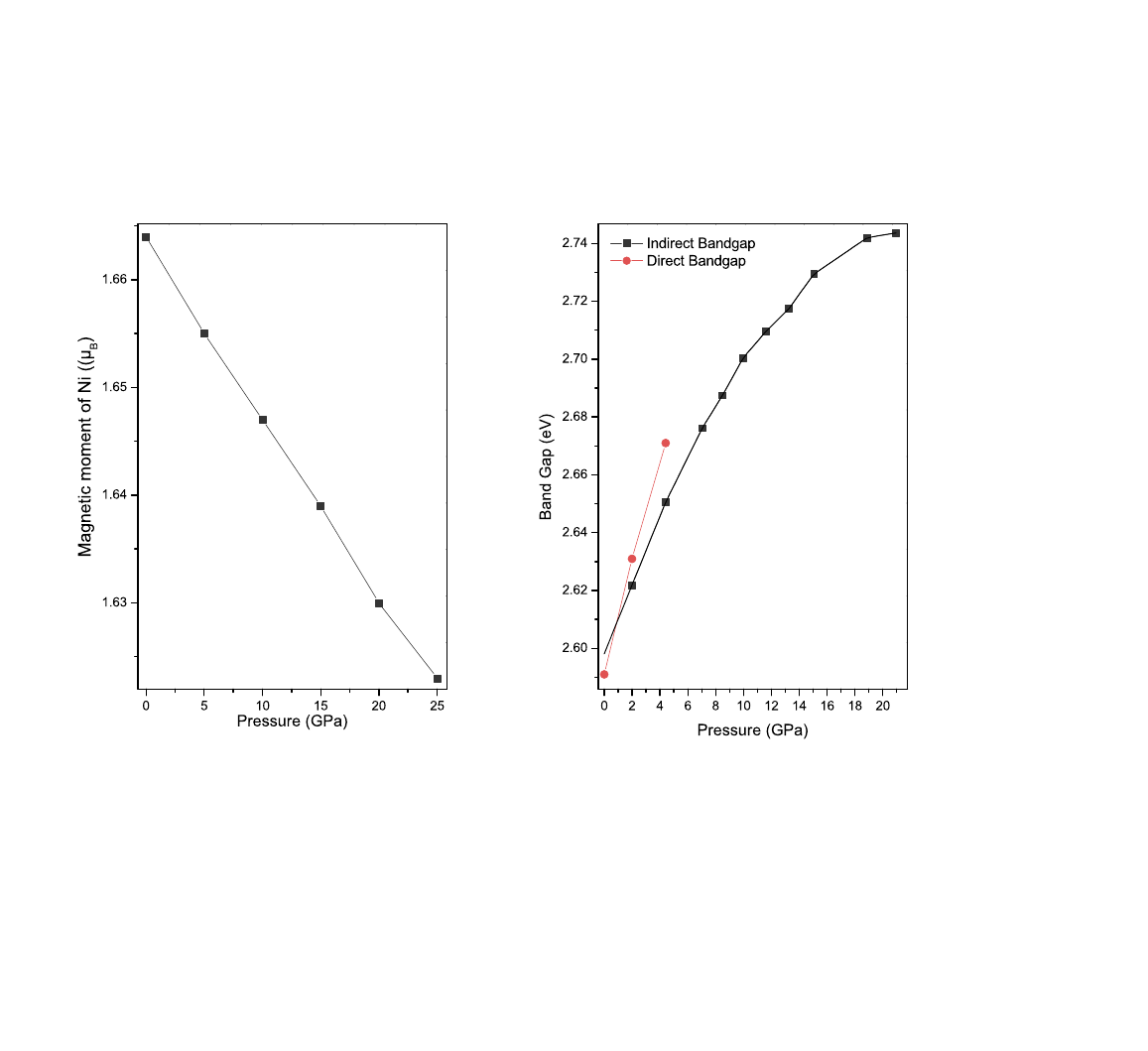}
\caption{(Color online) Computed pressure evolution of the magnetic moment per Ni site, in (a), and the bandgap energy, in (b). As it can be observed, there is a reduction of the magnetic moment potentially due to the Ni--O--Ni path modification, by the induced pressure, that alters the magnetic response.}
\label{fig:DFT_BG_MM}
\end{figure}

\end{document}